\documentclass[aps,showpacs,twocolumn]{revtex4}
\usepackage{epsfig}

\begin{document}

\title{Quantum chromodynamic quark benzene
\footnote{Email address: jlping@njnu.edu.cn (J. Ping)}}

\author{Jialun Ping$^{a,b}$, Chengrong Deng$^a$, Fan Wang$^c$, T. Goldman$^d$}

\affiliation{$^a$Department of Physics, Nanjing Normal University,
Nanjing 210097, P.R. China}

\affiliation{$^b$School of Physics and Microelectronics, Shandong
University, Jinan 280000, P.R. China}

\affiliation{$^c$Department of Physics, Nanjing University, Nanjing
210093, P.R. China}

\affiliation{$^d$Theoretical Division, Los Alamos National Laboratory, Los
Alamos, NM 87545, USA}

\begin{abstract}
A six-quark state with the benzene-like color structure based on a
color string model is proposed and studied. Calculation with
quadratic confinement with multi-string junctions shows that such a
state has a ground state energy similar to that of other hidden
color six-quark states proposed so far. Its possible effect on $NN$
scattering is discussed.
\end{abstract}



\maketitle

\bigskip

In the QED world, there are almost countless structures of matter.
Among organic molecules, there are very interesting varieties such
as methane, benzene, fullerene, etc. In the QCD world, up to now,
only very limited structures of matter have been discovered:
quark-antiquark mesons, three-quark baryons, nuclei and neutron
stars. However, the QCD interaction is richer and more varied than
QED so one would expect there to be even more varieties of QCD
matter. In pentaquark studies, various color structures have been
proposed: color singlet hadron molecules [K($q\bar{q}$)N($q^3$)],
color anti-triplet diquarks [$(qq)(qq)\bar{q}$], quark methane
[$q^4\bar{q}$], etc. Although the high statistics experimental data
of JLab and COSY~\cite{Jlab} did not reveal a pentaquark signal, the
theoretical view point of multi-quark structures has been broadened
in the course of pentaquark studies.

Lattice QCD calculation does not rule out the existence of
glueballs, quark-gluon hybrids, multi-quark states, etc. An
interesting question is whether there is a  peculiar property of low
energy non-perturbative QCD that is responsible for hiding or even
excluding these otherwise possible structures of matter. Lattice QCD
calculations for mesons, baryons, tetraquarks and pentaquarks reveal
a flux-tube or string like structure~\cite{lattice,latt1}. The naive
flux-tube or string model~\cite{isg,ww}, after adjusting the model
parameters, gives qualitatively similar estimates of ground state
energies of hadrons compared to other quark models. We therefore use
this model to study the possible structures of multi-quark systems.
Other models, such as the bag model~\cite{bag}, quark compound
model~\cite{simonov}, potential model~\cite{potential}, Skyrmion
model~\cite{kop}, chiral soliton model~\cite{dia} and
others~\cite{others} have all been used in the study of multi-quark
systems. In general they give similar results and mutually support
the existence of multi-quark systems.

In this letter, we propose and study a new color string structure, a
benzene-like structure, for the six-quark system. We estimate the
ground state energy of this system, which we call QCD benzene, using
a non-relativistic color string model~\cite{ww} with harmonic
confinement.

Strings are actually dynamical variables to which kinetic and
potential energies can be attributed~\cite{isg}. The potential
energy is proportional to the total length of strings. String
tensions can be estimated from the bag model or deduced from
lattice QCD calculations. They can also be determined empirically by
fitting hadron masses. We take the last, {\em i.e.} a purely
phenomenological approach. Since we are only interested in
qualitative properties of QCD benzene, we assume a quadratic
confinement potential for the color string rather than linear as this
simplifies the calculation significantly.

A comparative study showed the inaccuracy of this choice is quite
small~\cite{ww}. There are two reasons to expect this: One is that
the spatial variations in separation of the quarks (lengths of the
string) do not differ significantly, so the difference between the
two functional forms is small and can be absorbed in the adjustable
parameter, the stiffness. The second is that we are using a
non-relativistic description of the dynamics and, as was shown long
ago~\cite{shimon}, an interaction energy that varies linearly with
separation between fermions in a relativistic, first order
differential dynamics has a wide region in which a harmonic
approximation is valid for the second order (Feynman-Gell-Mann)
reduction of the equations of motion. A numerical check (see Table 1
below) confirmed this expectation.

For the six-quark system, there are four possible string structures
as shown in Fig.1. The first three were discussed in many papers in
the past but we have not found any previous discussion for the last
one. In these figures, $\mathbf{r}_i$ represents the position
coordinate of the quark $q_i$ which is denoted by a black dot,
$\mathbf{y}_i$ represents a junction where three flux tubes meet. A
thin line connecting a quark and a junction represents a
fundamental, {\em i.e.} color triplet, representation and a thick
line connecting two junctions is for a color sextet, octet or
others, namely a compound string. The different types of string may
have differing stiffness~\cite{mit,bali}. An inverted line
represents a conjugate $SU(3)$ color representation. Both the
overall color singlet nature of a quark system and the $SU(3)$ color
coupling rule at each junction must be satisfied.  Then various
types of color coupling are allowed in QCD, including that shown in
figure 1d of a benzene-like form.

\begin{center}
 \epsfxsize=1.8in \epsfbox{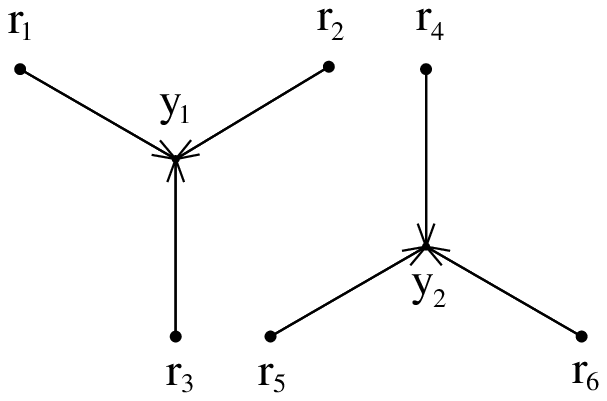}

  a. Two color-singlet baryons state

   \epsfxsize=2.3in \epsfbox{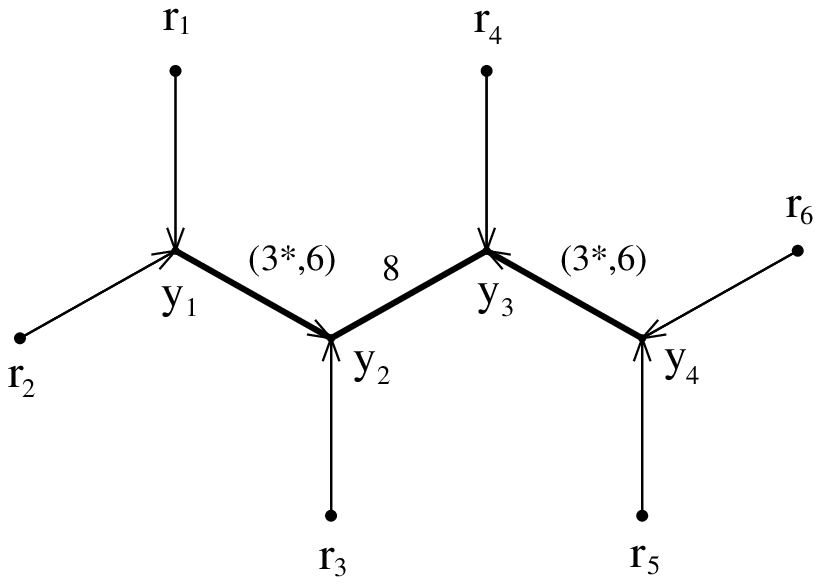}

 b. Two color-octet baryons state

 \epsfxsize=2.0in \epsfbox{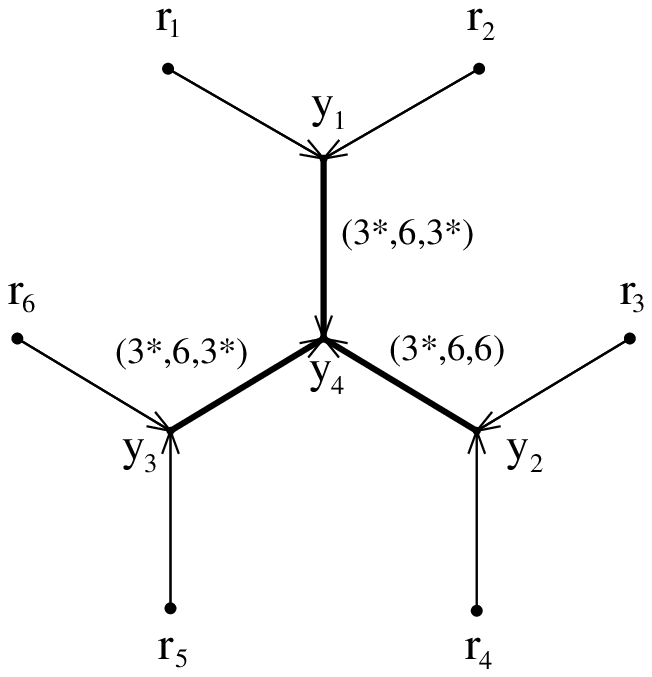}

  c. Three di-quarks state

 \epsfxsize=2.2in \epsfbox{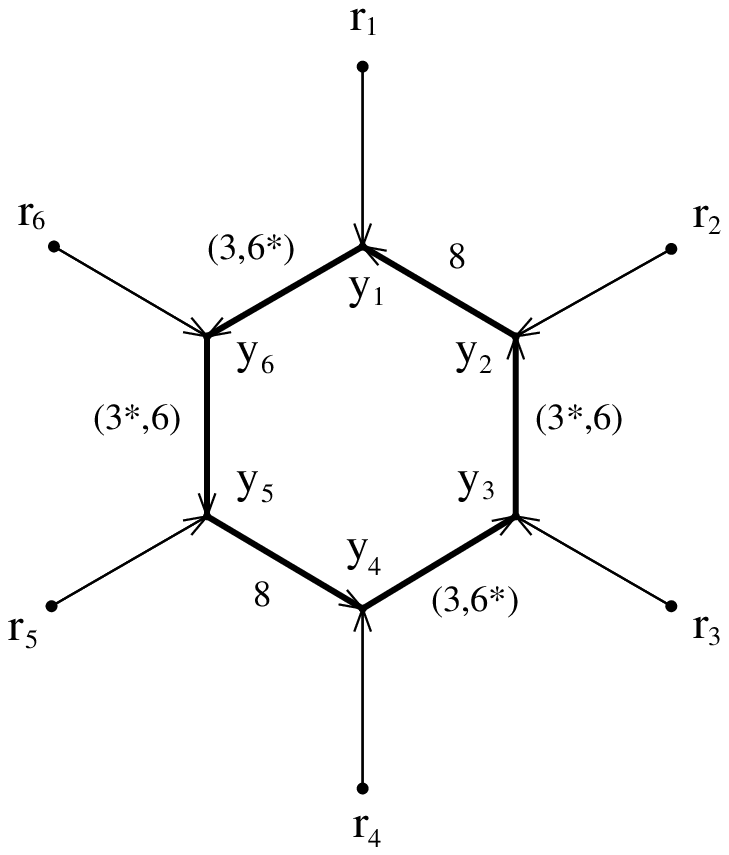}

 d. Benzene-like state

  Fig. 1. Four possible string structures of six-quark states.
\end{center}

In the string model with quadratic confinement, the potential energy
of the benzene-like string configuration for a six-quark state has
the following form,
\begin{equation}
U=\frac{1}{2}k\left( \sum_{i=1}^6(\mathbf{r}_{i}-\mathbf{y}_{i})^2
 + \kappa {\sum_{i<j}}^{\prime} (\mathbf{y}_i-\mathbf{y}_j)^2 \right),
\end{equation}
where the $\sum^{\prime}$ means the summation is over the adjacent
pairs. The string stiffness constant of an elementary or color
triplet string is $k$, while $k\kappa$ is the compound string
stiffness. In fig.1d, each quark is connected by a different
compound string and each string with a different dimension of the
color SU(3) group representation will have a different
stiffness~\cite{mit,bali}. However, due to quarks being fermions,
the Hamiltonian for a multi-quark system must be properly
antisymmetrized under quark exchange. In QED benzene, each carbon is
connected to the neighboring carbons with a single and a double
bond, but it has been proven that QED benzene has $D_{6h}$ symmetry.
Therefore, it is natural to use a common stiffness, $k\kappa$, for
the compound string for each color configuration, after taking into
account the symmetry requirement.

For given quark positions $\mathbf{r}_i,~i=1,\cdots,6$, in the
center-of-mass (CM) system, we can fix the position coordinates of
those junctions $\mathbf{y}_i,~i=1,\cdots,6$ by minimizing the
energy of the system. The result is,
\begin{equation}
 \mathbf{y}_1=\frac{(\kappa+4\kappa^2)(\mathbf{r}_2+\mathbf{r}_6)\\
 +\kappa^2(\mathbf{r}_3+\mathbf{r}_5)+(1+3\kappa)^2
 \mathbf{r}_1}{(1+\kappa)(1+3\kappa)(1+4\kappa)},
\end{equation}
and other configurations have a similar form. Thus we can define
a set of canonical coordinates,
\begin{eqnarray}
 \mathbf{R}_{1} & = & \frac{1}{2}(\mathbf{r}_6
  +\mathbf{r}_3-\mathbf{r}_1-\mathbf{r}_4), \nonumber \\
 \mathbf{R}_{2} & = & \frac{1}{2}(\mathbf{r}_6-\mathbf{r}_3
 +\mathbf{r}_1-\mathbf{r}_4), \nonumber \\
 \mathbf{R}_{3} & = & \sqrt{\frac{1}{12}}(2\mathbf{r}_2
 -\mathbf{r}_6-\mathbf{r}_1+2\mathbf{r}_5-\mathbf{r}_3-\mathbf{r}_4),
 \\
 \mathbf{R}_{4} & = & \sqrt{\frac{1}{12}}(2\mathbf{r}_2-\mathbf{r}_6
 +\mathbf{r}_1-2\mathbf{r}_5+\mathbf{r}_3-\mathbf{r}_4), \nonumber
 \\
\mathbf{R}_{5}
 & = & \sqrt{\frac{1}{6}}(\mathbf{r}_2+\mathbf{r}_4+\mathbf{r}_6
 -\mathbf{r}_1-\mathbf{r}_3-\mathbf{r}_5), \nonumber \\
 \mathbf{R}_{6} & = & \sqrt{\frac{1}{6}}(\mathbf{r}_1+\mathbf{r}_2
 +\mathbf{r}_3+\mathbf{r}_4+\mathbf{r}_5+\mathbf{r}_6). \nonumber
\end{eqnarray}
Finally, the potential energy $U$ can be written as,
\begin{eqnarray}
U & = & \frac{k}{2}\left(\frac{3\kappa}{1+3\kappa}\mathbf{R}_{1}^2
+\frac{\kappa}{1+\kappa}\mathbf{R}_{2}^2
+\frac{3\kappa}{1+3\kappa}\mathbf{R}_{3}^2 \right. \nonumber \\
&\ \ & \left. +\frac{\kappa}{1+\kappa}\mathbf{R}_{4}^2
+\frac{4\kappa}{1+4\kappa}\mathbf{R}_{5}^2\right)
\end{eqnarray}
Using this string potential in a non-relativistic Hamiltonian with
effective quark mass $m$~\cite{ww}, we find the ground-state
energy of QCD benzene to be
\begin{equation}
M_6=6m+\frac{3}{2}\omega\left(2\sqrt{\frac{\kappa}{1+\kappa}}
+2\sqrt{\frac{3\kappa}{1+3\kappa}}+\sqrt{\frac{4\kappa}{1+4\kappa}}
\right),
\end{equation}
where $\omega$ is equal to $\sqrt{k/m}$.

Furthermore, by calculating the distance between two adjacent quarks
and the distances between quarks and the CM, the spatial structure of
QCD benzene can be obtained. The distances between any two
adjacent quarks can be shown to be equal and so are the distances
between any quark and the CM. These are
\begin{eqnarray}
\langle (\mathbf{r}_i-\mathbf{r}_j)^2
\rangle&=&\frac{1}{2m\omega\sqrt{\kappa(1+4\kappa)}}\left(\sqrt{(1+\kappa)
(1+4\kappa)}\right.\nonumber \\
&\ \ & \left. +1+4\kappa+\sqrt{3(1+3\kappa)(1+4\kappa)} \right),
\end{eqnarray}
where $j = i+1$ (mod 6), and
\begin{eqnarray}
\langle {\mathbf{r}_i}^2
\rangle&=&\frac{1}{24m\omega\sqrt{\kappa(1+4\kappa)}}\left(12\sqrt{(1+\kappa)
(1+4\kappa)}\right.\nonumber \\
&\ \ & \left. +3+12\kappa+4\sqrt{3(1+3\kappa)(1+4\kappa)} \right).
\end{eqnarray}
Clearly these two distances generally are not equal, so the QCD
benzene cannot be planar.  However, we can show that the two
equilateral triangles, namely $\bigtriangleup_{135}$ and
$\bigtriangleup_{246}$, lie on two parallel planes. The length of
the side of the equilateral triangle is
\begin{equation}
\langle \mathbf{L}^2
\rangle=\frac{3}{2m\omega}\left(\sqrt{\frac{1+3\kappa}{3\kappa}}
+\sqrt{\frac{1+\kappa}{\kappa}}\right)
\end{equation}
and the distance between the two parallel planes is
\begin{equation}
\langle (\mathbf{R}_{135}-\mathbf{R}_{246})^2
\rangle=\frac{1}{2m\omega}\sqrt{\frac{1+4\kappa}{\kappa}}
\end{equation}
$\mathbf{R}_{135}$ and $\mathbf{R}_{246}$ represent the
centers-of-mass of $q_1q_3q_5$ and $q_2q_4q_6$, respectively. The
spatial structure of QCD benzene is shown in Fig.2. Projecting the
$\Delta_{135}$ plane onto the $\Delta_{246}$ plane, we obtain a
planar benzene.
\begin{center}
 \epsfxsize=1.9in \epsfbox{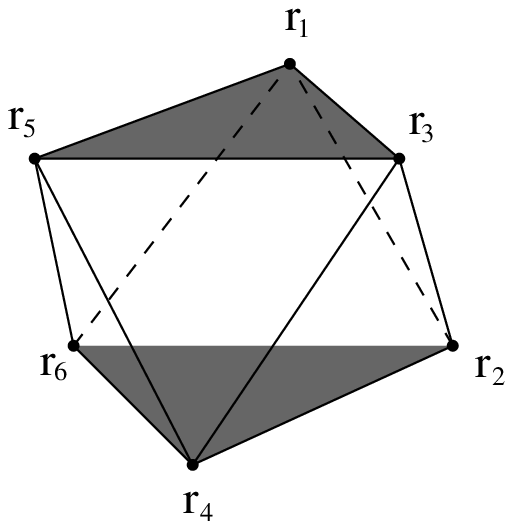}

 Fig. 2. The spatial structure of the benzene-like state.
\end{center}

To check this result, the same model has been applied to the tetraquark
and pentaquark systems. Three-dimensional configurations  are also
favored as shown in Figs.3 and 4, where a solid-dot and a hollow-dot
represent a quark and antiquark, respectively, while
$\mathbf{R}_{12}=\frac{\mathbf{r_1}+\mathbf{r_2}}{2}$ and
$\mathbf{R}_{34}=\frac{\mathbf{r_3}+\mathbf{r_4}}{2}$ represent the
midpoints of $\mathbf{r}_{1}$, $\mathbf{r}_{2}$ and
$\mathbf{r}_{3}$, $\mathbf{r}_{4}$, respectively. Please note Figs.3
and 4 only express the coordinates of the quarks; the string
configurations are not shown.

\begin{center}
 \epsfxsize=1.9in \epsfbox{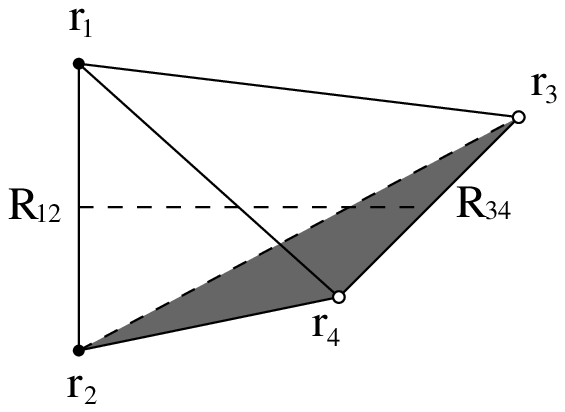}

Fig.3. The spatial structure of the tetraquark state.
\end{center}

In Figs.3 and 4, $\mathbf{r}_{12}=\mathbf{r}_1-\mathbf{r}_2$ is
perpendicular to $\mathbf{r}_{34}=\mathbf{r}_3-\mathbf{r}_4$, and
$\mathbf{r}_{12,34}=\mathbf{R}_{12}-\mathbf{R_{34}}$ is
perpendicular to $\mathbf{r}_{12}$ and $\mathbf{r}_{34}$. The
lengths of $\mathbf{r}_{12}$ and $\mathbf{r}_{34}$ can be shown to
be equal,
\begin{eqnarray}
\langle (\mathbf{r}_{12})^2 \rangle=\langle (\mathbf{r}_{34})^2
\rangle=\frac{1}{m\omega}
\end{eqnarray}
For the tetraquark and pentaquark states, the distances between
$\mathbf{R}_{12}$ and $\mathbf{R}_{34}$ are respectively:

\begin{center}
 \epsfxsize=1.9in \epsfbox{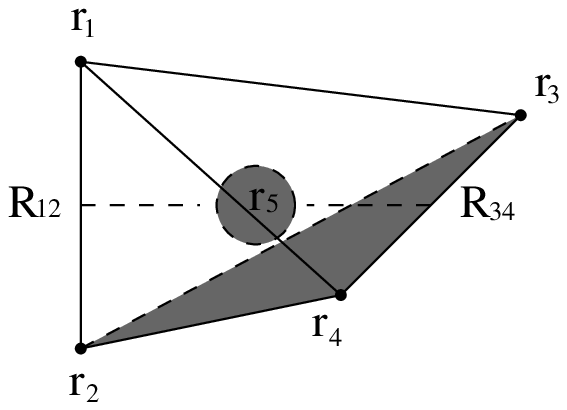}

Fig.4. The spatial structure of the pentaquark state.
\end{center}

\begin{eqnarray}
\langle (\mathbf{R}_{12}-\mathbf{R}_{34})^2
\rangle=\frac{1}{m\omega}\sqrt{\frac{1+\kappa}{\kappa}}
\end{eqnarray}
\begin{eqnarray}
\langle (\mathbf{R}_{12}-\mathbf{R}_{34})^2
\rangle=\frac{1}{m\omega}\sqrt{\frac{2+\kappa}{\kappa}}
\end{eqnarray}
In the pentaquark state, the distance between the antiquark and the
CM of the 4-quark subset is
\begin{eqnarray}
\langle (\mathbf{r}_{5,1234})^2
\rangle=\frac{1}{m\omega}\sqrt{\frac{2+5\kappa}{5\kappa}}
\end{eqnarray}
Consequently, multiquark states in our quark model must form
three-dimensional configurations which is due to the dynamics of the
systems: The string shrinks the distance between any two connected
quarks to as short a distance as possible to minimize the potential
energy, while the kinetic motion expands the distance between any
two quarks to as long a distance as possible to minimize the kinetic
energy: The stereo structures meet this requirement better than a
planar one does.

For pentaquark systems, lattice QCD obtains almost degenerate
energies for both planar and 3-dimensional configurations. However,
the entropy of a 5$q$ system is found to be larger in a
three-dimensional configuration~\cite{latt1}. For tetraquark
systems, a three-dimensional tetrahedral structure is rather stable
against the transition into two mesons~\cite{latt2}.

The non-zero value of $(\mathbf{r}_{5,1234})^2$ is due to the fact
that the anti-quark oscillates around the CM of the four quarks and
the value of $(\mathbf{r}_{5,1234})^2$ is calculated with a harmonic
oscillator wave function.

For comparison, we also present results for two other color structures
of the six quark system~\cite{ww},
\begin{eqnarray}
M_{33}&=&6m+\frac{3}{2}\omega\left(2+\sqrt{\frac{3\kappa}{2+3\kappa}}
+\sqrt{\frac{\kappa(2\kappa+0.6277)}{2\kappa^2+7\kappa+2}} \right.
\nonumber \\
&\ \ & \left.
+\sqrt{\frac{2\kappa(\kappa+3.186)}{2\kappa^2+7\kappa+2}} \right)
\end{eqnarray}
\begin{equation}
M_{222}=6m+\frac{3}{2}\omega\left(3+2\sqrt{\frac{\kappa}{2+\kappa}}\right)
\end{equation}
Here, $M_{33}$ and $M_{222}$ are ground-state energies of the two color-octet,
hidden color state (Fig.1b) and the three di-quark state (Fig.1c) respectively.

In our calculation, there are three free parameters: The non-strange quark
mass $m$ and the harmonic oscillator stiffnesses, $k$ and $\kappa k$. The
first two can be fixed from the masses of the non-strange hadrons. We
take $\overline{M}_{3}$ as the average mass of $N+\Delta$ and take
$\overline{M}_{2}$ as the average mass of light, non-strange mesons
\begin{eqnarray}
 \overline{M}_{3} & = & \frac{1}{2}(N+\Delta) = 3m+3\omega, \\
 \overline{M}_{2} & = & \frac{1}{4}\left(\frac{3}{4}\pi+
\frac{1}{4}\eta\right)+ \frac{3}{4}\left(\frac{3}{4}\rho+\frac{1}{4}\omega
\right) \nonumber \\
 & = & 2m+\frac{3}{2}\omega.
\end{eqnarray}
From this, we obtain the effective quark mass
\begin{equation}
 m = 2\overline{M}_{2}-\overline{M}_{3} = 0.19 GeV.
\end{equation}
The constant $k$ of an elementary string can be eliminated in favor
of the experimental baryon mass $\overline{M}_{3}$. The
inter-junction string parameter $\kappa$ depends on the color
dimension, $d$, of the string. In the MIT bag model~\cite{mit}, for
example, one finds effectively
\begin{equation}
 \kappa_{d}=\sqrt{\frac{C_d}{C_3}},
\end{equation}
where $C_d$ is the eigenvalue of the Casimir operator associated
with the $SU(3)$ color representation $d$ on either end of the
string. However lattice QCD calculations find that the stiffness,
$\kappa_d$, of higher dimensional color charge is scaled by $C_d$
instead of $\sqrt{C_d}$~\cite{bali}. The compound strings of
six-quark systems have five possible color structures: $\textbf{3}$,
$\textbf{3*}$, $\textbf{6}$, $\textbf{6*}$ and $\textbf{8}$, so each
geometric configuration may have different states when the color
dimension dependence of the string stiffness is included. This means
that the minimum string energy also varies in general from one state
to another. We shall not include all these (small) variations, but
use instead an average compound string constant $\kappa k$.

Numerical results for these three hidden color structures are shown
in the Table 1. To check the approximation introduced by quadratic
confinement the linear confinement results with the same multi-quark
wave functions obtained from quadratic confinement model are listed
in Table 1 as a comparison.
\begin{center}
Table 1. The masses of the six-quark systems (GeV)

\begin{tabular}{|c|c|c|c|c|}
\hline
 confinement & $\kappa$ & Three-di-quark &Color-octet& Benzene-like
 \\ \hline
  quadratic &1.0& 2.22 & 2.21&2.19\\ \cline{2-5}
   &1.5& 2.26 & 2.26 & 2.25\\ \cline{1-5}
 linear & 1.0 & 2.31 & 2.31 & 2.38 \\ \cline{2-5}
  & 1.5 & 2.31 & 2.31 &2.37 \\ \hline
\end{tabular}
\end{center}

From the Table, it can be seen that while the different color
structures have similar masses (around 2.2 to 2.4 GeV), the
benzene-like structure has the lowest mass whereas the three
di-quark structure has the highest mass for quadratic confinement.
On the other hand for linear confinement, the benzene-like structure
has the highest mass, whereas the other two have smaller masses.
After taking into account the model uncertainties, we can only
conclude that the three hidden color states have similar masses.
With an increase of $\kappa$, the differences between the masses
decreases. In the limit $\kappa\rightarrow\infty$, they have the
same mass due to having the same flux-tube structure in which the
compound flux-tube shrinks to zero, leaving a hub and spokes
configuration with one junction.

The color magnetic interaction (CMI), which is important for hadron
spectroscopy, has not yet been included in our model calculation.
The main reason for this is that we have not found a method to
calculate the matrix element of CMI for benzene-like structure. In
our previous pentaquark calculation, we found that the tetrahedron
has a lower mass than the Jaffe-Wilczek di-quark structure even
after taking into account of CMI~\cite{ping}.
To give a estimate of the effect of CMI, we calculated CMI for
hidden color states consisted of two color-octet baryons. the
results are in the range of 20-120 MeV for two color-octet nucleons
or $\Delta$'s systems. We expect the benzene-like structure should
have energy around 2.21-2.50 GeV, at 0.33-0.62 GeV above the
two-nucleon mass.

The true six quark state will, of course, involve a mixture of these
differing string structures.
In general, channel coupling will reduce the ground state energy of
the system. Therefore,
for channel coupling calculations in QCD models, these color
channels should be included. Unfortunately, at present  we do not
have any reliable information about the transition interaction
between different color structures, especially from a color singlet
hadron state to genuine hidden color multi-quark states. The
transition interaction is very much needed for the study of
multi-quark states.

In nucleon-nucleon ($NN$) scattering, QCD benzene should be one of
the possible intermediate states as well as the other color
structures shown in Fig.1. In order to simplify our argument, we
neglect the 6 and 6* strings temporarily. Then, if we cut the two
color 8 strings, the QCD benzene will shrink into two color singlet
baryons (which we take to be two nucleons for convenience in this
discussion). Except for exchanges in quark labels, these two
nucleons are the initial ones, as shown in Fig.1a, in the $NN$
scattering process.
    It should be possible for two nucleons to transform into
QCD benzene and other genuine hidden color six quark states due to
flux tube fluctuations when they are close enough to be
within the range of confinement ($\sim 1 fm$). For example, the
creation of two gluons can simultaneously excite the two color singlet
nucleons into two color octet baryons which are coupled together to
be overall color singlet  -- this is nothing else but a QCD benzene.
    If the scattering energy is near that of these hidden color states
(2.2 to 2.5 GeV), then the intermediate six quark state should be
dominantly in a hidden color state because the number of hidden
color states is much larger than color singlet one so the entropy of
hidden color state is much higher than the color singlet one and it
cannot decay into two colorful nucleons directly due to color
confinement. It must transform back into two color singlet nucleons
before decaying. Such a process is similar to compound nucleus
formation and therefore should induce a resonance around the
energies of these hidden color states in $NN$ scattering. We call
this state a ``color confined, multi-quark resonance"  and it is
different from all of the microscopic resonances discussed by S.
Weinberg~\cite{weinberg}. Our present understanding of QCD does not
obviate such a back and forth transition between color singlet
hadron and genuine hidden color multi-quark states. An open question
remains as to whether or not experiments have observed such a color
confined, multi-quark resonance. In fact there is a structure in the
$pp$ scattering cross section around 2.2 to 2.4 GeV which has been
explained in terms of $\Delta$ production~\cite{mach}.
However in Our quark model calculations, we found that there are
several $N-\Delta$, $\Delta-\Delta$ broad resonances in this energy
range, in addition to the benzene-like structure. Can it also be
described by broad overlapping resonances due to these structures?
This possibility seems not to have been ruled out and it is under
study in our group.

The possible existence of benzene-like structure is based on the
description of quarks interaction via flux tubes. Although the
lattice QCD calculation gave some supports to the idea, its validity
is not beyond doubt. Other studies are needed to check if the QCD
benzene structure obtained from color flux tube model is really a
QCD result.

 \vspace{8mm}

This work is supported by the National Science Foundation of China
under grants 10375030 and 90503011 and by the U.S. Department of
Energy under contract W-7405-ENG-36.

\end{document}